%% file: main.tex
\documentclass[runningheads]{llncs}
\usepackage[T1]{fontenc}
\input{src/tex/pre}
\usepackage{graphicx}
\usepackage{todonotes}
\begin{document}
\title{\candy{} SuiteEval: Simplifying Retrieval Benchmarks}

\author{Andrew Parry\inst{1}\orcidID{0000-0001-5446-8328} \and
Debasis Ganguly\inst{1}\orcidID{0000-0003-0050-7138} \and
Sean MacAvaney\inst{1}\orcidID{0000-0002-8914-2659}}

\institute{University of Glasgow, Glasgow, Scotland, UK\\
\email{a.parry.1@research.gla.ac.uk} \\ \email{\{Debasis.Ganguly, Sean.MacAvaney\}@glasgow.ac.uk}}
\maketitle              %
\begin{abstract}
Information retrieval evaluation often suffers from fragmented practices---varying dataset subsets, aggregation methods, and pipeline configurations---that undermine reproducibility and comparability, especially for foundation embedding models requiring robust out-of-domain performance. We introduce SuiteEval, a unified framework that offers automatic end-to-end evaluation, dynamic indexing that reuses on-disk indices to minimise disk usage, and built-in support for major benchmarks (BEIR, LoTTE, MS MARCO, NanoBEIR, and BRIGHT). Users only need to supply a pipeline generator. SuiteEval handles data loading, indexing, ranking, metric computation, and result aggregation. New benchmark suites can be added in a single line. SuiteEval reduces boilerplate and standardises evaluations to facilitate reproducible IR research, as a broader benchmark set is increasingly required.

\vspace{0.6em}
\hspace{5em}\includegraphics[width=1.25em,height=1.25em]{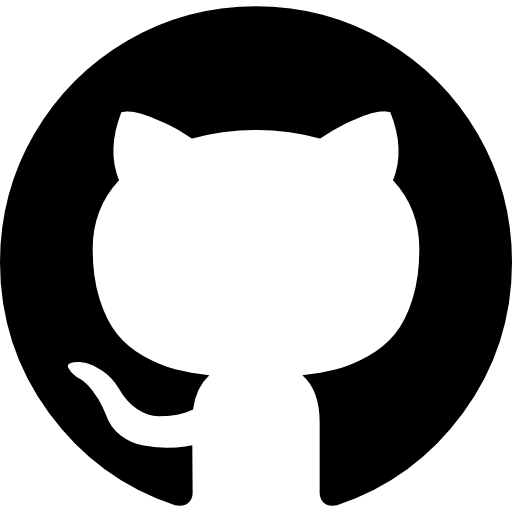}\hspace{.3em}
\parbox[c]{\columnwidth}
{
    \vspace{-.55em}
    \href{https://github.com/Parry-Parry/suiteeval}{\nolinkurl{https://github.com/Parry-Parry/suiteeval}}
}
\vspace{-1.2em}
\end{abstract}
\section{Introduction}

\input{src/tex/intro}

\section{\candy $ \ \text{SuiteEval}$ Overview}
\input{src/tex/components}

\section{Demonstration}
\input{src/tex/demonstration}

\section{Target Audience and Potential Use Cases}
\input{src/tex/usage}
\section*{Disclosure of Interests}
The authors have no competing interests to declare that are relevant to the content of this article.

\bibliographystyle{splncs04}
\bibliography{bib}
\end{document}

%% file: src/tex/pre.tex
\newcommand{\candy}{%
  \includegraphics[height=1em]{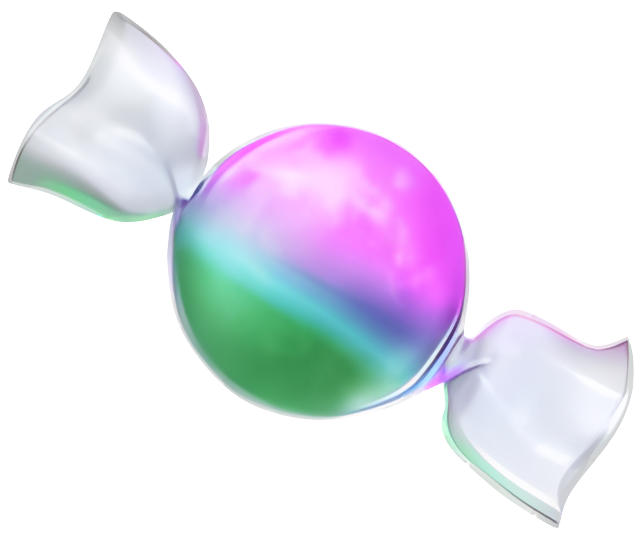}%
}
\usepackage{comment}

\usepackage{listings}
\usepackage{xcolor}
\usepackage{flushend}
\usepackage{multicol}
\usepackage{booktabs}
\usepackage{hyperref}
\usepackage{algorithm}
\usepackage{amsmath}
\usepackage{algpseudocode}
\lstdefinestyle{mystyle}{
    language=Python,
    basicstyle=\ttfamily\footnotesize,
    keywordstyle=\color{blue}\bfseries,
    commentstyle=\color{gray}\itshape,
    stringstyle=\color{red},
    numberstyle=\tiny\color{gray},
    numbers=left,
    stepnumber=1,
    numbersep=5pt,
    backgroundcolor=\color{lightgray!20},
    frame=single,
    rulecolor=\color{black},
    breaklines=true,
    breakatwhitespace=true,
    captionpos=b,
    tabsize=4,
    showspaces=false,
    xleftmargin=10pt,    %
    aboveskip=5pt,      %
    belowskip=5pt,      %
    columns=flexible,    %
}

%% file: src/tex/intro.tex
Evaluation benchmarks are central to measuring effectiveness in information retrieval (IR). As the community increasingly targets out-of-domain generalisation---motivated by ``foundation'' embedding models intended to operate across diverse domains and surface forms~\cite{contriever,retromae}---studies are expected to report results over multiple collections and splits, often bundled into a \textit{suite} of benchmarks. In practice, however, evaluations are frequently constrained by the systems natively supported by a suite's accompanying codebase, which can lag behind new architectures. Further, choices about dataset coverage and result aggregation frequently differ across studies~\cite{tailoftwo}, complicating comparison and replication.

Open-source toolkits such as \texttt{PyTerrier}~\cite{pyterrier} and \texttt{Anserini}~\cite{anserini} have lowered the barrier to reproducible experimentation by standardising data ingestion and facilitating pipeline construction. Yet they provide limited support for managing end-to-end suite evaluations comprising \emph{multiple} benchmarks: enforcing consistent dataset selection, unifying per-benchmark metric definitions, automating indexing and ranking across all splits, and reducing the incidental engineering that surrounds large-scale runs. Furthermore, while toolkits such as \texttt{ir\_datasets}~\cite{irds} provide consistent data processing and \texttt{Tira}~\cite{tira} facilitates reproducibility through containerization, we instead target the convenient and memory-efficient evaluation of end-to-end pipelines over multiple benchmarks.

This work introduces \candy $ \ \texttt{SuiteEval}$, a unifying framework that simplifies and standardises suite evaluation in IR. \texttt{SuiteEval} encapsulates benchmark definitions, dataset contexts, indexing, and ranking within a single \emph{Suite} abstraction that: (i) defaults to complete coverage of all datasets and splits for a benchmark suite; (ii) applies the official metrics and aggregation procedures uniformly; and (iii) eliminates boilerplate orchestration so that users specify only the IR pipeline to be evaluated. A dynamic indexing process minimises superfluous disk usage over large suites while preserving reproducibility and auditability of artefacts. By standardising evaluation over widely used suites, we hope to reduce the burden on performing suite-based evaluation, while reducing the variance in experimental settings and making effectiveness claims more comparable across papers.

We demonstrate \texttt{SuiteEval} by executing a standard experiment over BEIR and the creation of a new benchmark, guiding users through the minimal steps required to execute reproducible benchmark suite evaluations. Together, these results show how \texttt{SuiteEval} streamlines reproducible IR evaluation while enabling straightforward extension as new suites are released.

%% file: src/tex/components.tex
Many IR evaluation setups are driven by declarative configuration files that fix index paths, dataset splits, and pipeline parameters. This encourages duplicated indices per corpus and boilerplate for each new retrieval or re-ranking component, inflating disk usage and increasing variance in evaluation settings. Here we adopt \texttt{PyTerrier} pipelines for their composability and broad existing ecosystem: pipeline operators (e.g., \texttt{$\gg$}) make retrieval-re-ranking stages explicit within a single, minimal specification.
\vspace{1.5em}

\noindent\textbf{Running an Evaluation Suite.}
Existing experimental procedures assume that experimental artifacts (e.g., indexes) are constructed before evaluation. This poses a challenge for evaluation suites, where many potential artifacts are required and may only be used once. \texttt{SuiteEval} overcomes this by defining the process for constructing required artifacts and retrieval pipelines in a single user-defined function (example in Figure \ref{fig:systems}). As input to this function, each dataset is wrapped in a \texttt{DatasetContext}, which provides (i) a workspace via \texttt{path}, (ii) a corpus iterator via \texttt{get\_corpus\_iter()} (required for constructing indexes), and (iii) dataset-specific utilities such as \texttt{text\_loader()}. SuiteEval is responsible for executing and evaluating these pipelines.
\begin{figure}
\begin{lstlisting}[language=Python]
from suiteeval import DatasetContext
from pyterrier_pisa import PisaIndex
import pyterrier_t5, pyterrier_dr
def systems(context: DatasetContext):
    index = PisaIndex(context.path / "index.pisa")
    index.index(context.get_corpus_iter())
    bm25 = index.bm25() >> context.text_loader()
    yield bm25 >> pyterrier_t5.MonoT5ReRanker()
    yield bm25 >> pyterrier_dr.ElectraScorer()
\end{lstlisting}
\caption{Creation of a \texttt{PisaIndex} using BM25 before re-ranking with two neural models.}
    \label{fig:systems}
\end{figure}
All filesystem writes are scoped to \texttt{context.path}, treated as ephemeral unless a persistent \texttt{index\_dir} is supplied. Indices (e.g., \texttt{PisaIndex}) are instantiated inside this workspace and reused if present; otherwise they are built once per corpus using \texttt{context.get\_corpus\_iter()}. Because \texttt{systems(context)} is invoked once per corpus (grouping multiple query sets together that use the same corpus), all yielded pipelines share the same index instance, and \texttt{text\_loader()} attaches document text only when required by downstream re-rankers.

\noindent \textbf{The \texttt{Suite} Class.}
A \texttt{Suite} instance acts as the controller, binding dataset identifiers, official measures, and aggregation rules. The suite groups dataset identifiers by underlying corpus so that indexing is performed once per corpus rather than per test collection. For each corpus group, it instantiates a \texttt{DatasetContext}, invokes the user generator once to collect pipelines, and executes those pipelines across all associated test collections. A minimal invocation is shown in Figure \ref{fig:beir}.
\begin{figure}
    \begin{lstlisting}[language=Python]
from suiteeval import BEIR

results = BEIR(systems, baseline=0, save_dir="beir_results")
\end{lstlisting}
    \caption{Execution of the BEIR evaluation suite. MonoT5 is taken to be the baseline for significance tests and run files are saved to ``beir\_results''.}
    \label{fig:beir}
\end{figure}
Results are returned as a long-form \texttt{pandas.DataFrame}, with one row per dataset, system, and evaluation measure. Because all datasets in a suite are evaluated under a single controller with fixed measures and cutoffs, SuiteEval can additionally compute suite-level aggregates, such as means across datasets. Existing suites predefined within SuiteEval are listed in Table~\ref{tab:suiteeval-benchmarks}.

\noindent \textbf{Adding New Suites.}
\texttt{SuiteEval} provides a registry mechanism that allows new benchmarks to be defined declaratively without subclassing or custom evaluation scripts. The \texttt{Suite.register} method associates a suite name with a list of \texttt{ir\_datasets} identifiers and a metadata dictionary, as shown in Figure \ref{fig:define}. This metadata configures evaluation behaviour, including the set of official measures and default aggregation rules. Once registered, suites are first-class objects and can be invoked in the same manner as built-in benchmarks.
\begin{figure}
    \begin{lstlisting}[language=Python]
from suiteeval.suite.base import Suite
from ir_measures import nDCG

passage_datasets = [
    "msmarco-passage/trec-dl-2019/judged",
    "msmarco-passage/trec-dl-2020/judged",
    "msmarco-passage-v2/trec-dl-2022/judged"]

MSMARCOPassage = Suite.register(
    "msmarco/passage",
    datasets=passage_datasets,
    metadata={"official_measures": [nDCG@10]}
    )
\end{lstlisting}
    \caption{Definition of a custom suite comprising two corpora (MSMARCOv1 and MSMARCOv2) and 3 test collections (DL-2019, -2020, -2022). This suite will return nDCG@10 values for each test collection and the geometric mean of the three for each pipeline defined in the \texttt{systems} function.}
    \label{fig:define}
\end{figure}

At runtime, the registered suite resolves each dataset identifier into its corresponding corpus, topics, and relevance judgements using \texttt{ir\_datasets}\footnote{Other objects may be used, but they must follow the \texttt{ir\_datasets} object structure.}. The declared evaluation measures are applied uniformly across all datasets, ensuring that metric definitions cannot drift between splits. Because suite registration occurs at import time, adding new benchmarks introduces no runtime overhead and does not require changes to experimental scripts.

\begin{table*}[t]
\centering
\small
\setlength{\tabcolsep}{2.5pt}
\caption{Pre-registered suites currently supported by \candy \texttt{SuiteEval}.}

\label{tab:suiteeval-benchmarks}
\begin{tabular}{@{}lp{8.5cm}@{}}
\toprule
\textbf{Name} & \textbf{Scope} \\
\midrule

 \texttt{Lotte()}~\cite{lotte} & Long-tail collections. \\[2pt]

\texttt{MSMARCOPassage()}~\cite{msmarco} & MS~MARCO passage tasks~\cite{dl2019,dl2020}. \\[2pt]

\texttt{MSMARCODocument()}~\cite{msmarco} & MS~MARCO document tasks~\cite{dl2019,dl2020}; mirrors passage-suite behaviour at document level. \\[2pt]

\texttt{BEIR()}~\cite{beir} & Heterogeneous, multi-domain retrieval suite; additionally applies post-hoc filtering and corrects aggregation. \\[2pt]

\texttt{NanoBEIR()} & Compact BEIR subsets for fast iteration. Similarly post-processing to BEIR. \\[2pt]

\texttt{BRIGHT()}~\cite{bright} & Collections targeting reasoning intensive tasks. \\
\bottomrule
\end{tabular}
\end{table*}

\noindent \textbf{Reducing Memory Footprint.}
\texttt{SuiteEval} explicitly bounds index materialisation by the number of distinct corpora rather than by the number of test collections or pipeline variants. A single index persists only for the duration of a corpus group and is reused across all associated test collections before being released. For example, when evaluating the MS~MARCO passage suite, an MSMARCOv1 index is reused for DL-2019 and DL-2020 before being destroyed and replaced by an MSMARCOv2 index for DL-2022. This lifecycle underpins the storage reductions reported in Table~\ref{tab:suiteeval-experiment}.

\begin{table*}[t]
\centering
\small
\setlength{\tabcolsep}{2pt}
\caption{Storage footprint of common experiments with and without \candy $ \ \texttt{SuiteEval}$. The bi-encoder applied end-to-end and to re-rank BM25 is RetroMAE.}
\label{tab:suiteeval-experiment}
\begin{tabular}{@{}llrr@{}}
\toprule
\textbf{Experiment} & \textbf{Suite} & \textbf{Disk Space (MB)} & + \candy $ \ \textbf{SuiteEval}$ \\
\midrule
End-to-End vs. Re-Ranker & NanoBEIR & 249.85 & \textbf{18.29} \\
BM25 Grid-Search & BEIR & 22888.07 & \textbf{4889.15} \\
\bottomrule
\end{tabular}
\end{table*}

%% file: src/tex/demonstration.tex
Our demonstration presents a concrete application of our framework, starting with a simple demonstration of running a grid search over a large benchmark suite. It explains what is provided within a suite and can be explored in this \href{https://drive.google.com/file/d/1NdbMxaUZw6znL6lde9nTsMelr23MsfgJ/view?usp=sharing}{live notebook}. We then provide a walkthrough on how to create a simple suite and extend its functionality to be task-specific by applying custom aggregation to results, and it can be explored in this \href{https://drive.google.com/file/d/1HjTIoSj6RKHnyGibKkc7n-wIhaXNccf2/view?usp=sharing}{live notebook}.  Additionally, we aim to raise awareness of the need for consistent and reproducible evaluation, a concept often championed in the IR community, with our framework serving as a tool for the responsible use of benchmark suites.

%% file: src/tex/usage.tex
The target audience for this demo is PhD students and researchers working in neural information retrieval or, more broadly, in representation learning with a focus on document embeddings. By introducing our package and utilising the broader PyTerrier ecosystem within it, we aim to continue encouraging reproducible evaluation, as benchmark suites become increasingly common as primary indicators of effectiveness.